\documentclass[aps,prl,showpacs,twocolumn]{revtex4}
\usepackage[dvips]{graphicx,color}
\usepackage{bm,amsmath}
\begin{document}
\title{Friction in nanoelectromechanical systems: Clamping loss in the GHz regime}
\author{Michael R. Geller$^1$ and Joel B. Varley$^2$}
\affiliation{$^1$Department of Physics and Astronomy, University of Georgia, Athens, Georgia 30602-2451 \\ $^2$Department of Physics, University of California, Santa Barbara, California 93106}

\date{December 29, 2005}

\begin{abstract}
The performance of a wide variety of ultra-sensitive devices employing nanoelectromechanical resonators is determined by their mechanical quality factor, yet energy dissipation in these systems remains poorly understood. Here we develop a comprehensive theory of friction in high frequency resonators caused by the radiation of elastic energy into the support substrate, referred to as clamping loss. The elastic radiation rate is found to be a strong increasing function of resonator frequency, and we argue that this mechanism will play an important role in future microwave-frequency devices.
\end{abstract}

\maketitle

Nanoelectromechanical systems (NEMS) have enormous potential for both fundamental research and novel device applications \cite{Craighead00,RoukesPW01,BlickJPCM02,Blencowe04,EkinciRSI05}. High-frequency NEMS resonators are the leading candidates in a current effort to observe quantum behavior in a macroscopic mechanical system \cite{KnobelNat03,LaHayeSci04,SchwabPT05}. Achieving the quantum limit requires resonator frequencies $\omega$ that are significantly higher than $k_{\rm B} T/\hbar$, as well as high mechanical quality factors
\begin{equation}
Q \equiv \omega \tau,
\end{equation} 
where $\tau$ the energy relaxation time. In this limit the resonator's phonons will be analogous to photons in an electromagnetic cavity, and a variety of mechanical quantum ``optics" effects have been discussed in the literature, including lasing \cite{BargatinPRL03}, squeezing \cite{Blencowe00,RabiPRB04,RuskovPRB05}, cooling \cite{MartinPRB04,ZhangPRL05}, and quantum nondemolition measurement \cite{SantamorePRB04}. Macroscopic quantum tunneling of a low-temperature beam has been predicted \cite{CarrPRB01,WernerEL04}, and methods to entangle NEMS with other quantum systems have been proposed \cite{ArmourPRL02,MarshallPRL03,IrishPRB03}, which may lead to their use in quantum computing \cite{Cleland&GellerPRL04,ZouPLA04,Geller&ClelandPRA05,Pritchett&GellerPRA05,SunPre05}. NEMS operating in the classical regime have already been used successfully in several novel devices, particularly for sensing and actuation. They have been used to study the Casimir force at short distances \cite{LamoreauxPRL97,MohideenPRL98}, and can be used as ultra-sensitive mass sensors \cite{OnoRSI03,EkinciJAP04}, recently detecting mass changes with zeptogram sensitivity \cite{roukesnote}. They have demonstrated enormous potential in scanning probe microscopy \cite{SidlesRMP95}, including the ability to detect the spin of a single electron \cite{RugarNat04}, and may also find future applications in classical signal processing \cite{BuksJMEMS02,LifshitzPRB03,BadzeyAPL04,CrossPRL04,BadzeyNat05}.

The performance of a mechanical resonator improves with increasing $Q$. However, in a series of experiments on small mechanical resonators, performed over many years by several different groups, the quality factor for motion in the fundamental flexural and other low-lying vibrational modes has been observed to decrease approximately linearly with decreasing resonator size \cite{MihailovichPRL92,GreywallEL96,BiggarPRB97,CarrAPL99,EvoyJAP99,EvoyAPL00,YangAPL00,YasumuraJMEMS00,SpielRSI01,LiuAPL01,MohantyPRB02,YangJMEMS02}. This dependence is usually interpreted to be a signature of surface contamination effects, such as dissipation caused by adsorbates or surface reconstructions. However, measurements of the temperature dependence of $Q$ also suggest that glass-like structural two-level systems \cite{KleinmanPRL87,PhillipsPRL88,KeyesPRL89,AhnPRL03} or other internal ``bulk'' lattice \cite{GreywallEL96} or electronic \cite{BiggarPRB97,MohantyPRB02} defects may also play an important role. Many energy dissipation mechanisms have been proposed to explain these experiments. However, most explanations are qualitative in nature, and a detailed quantitative analysis of the competing mechanisms has not yet been carried out. 

\begin{figure}
\includegraphics[width=9.5cm]{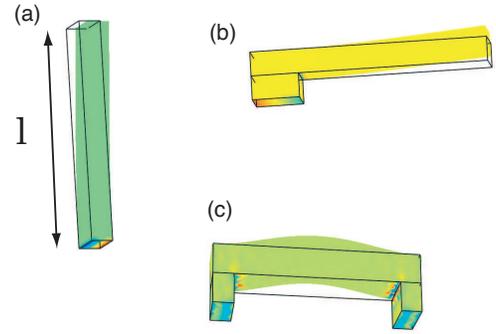}
\caption{(Color online) Simulated resonators. (a) $2.4 \, {\rm GHz}$ perpendicular beam resonator. (b) $3.2 \, {\rm GHz}$ cantilever. (c) $16 \, {\rm GHz}$ doubly clamped beam. The material is Si and $l \! = \! 50 \, {\rm nm}$ in the examples shown. The surfaces are colored according to the $z$ component of their traction, with yellow representing zero. The lower surfaces are attached to a semi-infinite elastic substrate.}
\label{cantilever figure}
\end{figure} 

An obvious mechanism for energy dissipation in a mechanical resonator is elastic radiation loss through the resonator supports, where the resonator is attached to a bulk substrate \cite{Jimbo68,CrossPRB01,PhotiadisAPL04}. Any mechanical interaction of the resonator with its surroundings will allow vibrational energy to escape to the substrate. But little is understood quantitatively about this so-called clamping loss mechanism, and some existing estimates are in contradiction with experiment. For example, the estimate \cite{CrossPRB01} 
\begin{equation}
Q_{\rm CL} \approx 0.3 \bigg(\frac{l}{w}\bigg)^{ \! 3}
\end{equation}  
for an in-plane flexural mode applied to the $1 \, {\rm GHz}$ SiC beam of Huang {\it et al.}~\cite{HuangNat03} results in a $Q$ factor of about 230,  smaller than observed. In the experiment of Ref.~[\onlinecite{HuangNat03}], the beam length is $l \! = \! 1.1 \, {\rm \mu m}$ and width is $w \! = \! 120 \, {\rm nm}$ \cite{dimensionsnote}. 

In this paper we develop a general method for the accurate calculation of clamping loss rates for a wide variety of NEMS geometries. We assume at the outset that the clamping loss is small (its contribution to the total $Q^{-1}$ much less than unity), allowing a perturbative treatment. If the total dissipation rate $Q^{-1}$ is also $\ll 1$, as in the case of experimental interest, it will be given by a sum of contributions from each dissipation channel present. Our results suggest that clamping loss will play an important role in future GHz-frequency devices, especially above $10 \, {\rm GHz}$.

We begin our analysis with a discussion of the fundamental mode \cite{fundamentalnote} of the perpendicular beam resonator illustrated in Fig.~\ref{cantilever figure}a, with frequency $\omega$. The lower surface is to be attached to a semi-infinite elastic substrate, and all exposed surfaces are stress free. $\omega$ will always be degenerate with vibrational modes of the substrate, so how can clamping loss ever be small? The answer is that an elastic wave present in the beam will be strongly reflected at the substrate when its emitted wavelength $\omega/v$, which for the fundamental mode is of the order of $l$, is larger than the transverse dimensions of the beam. Here $v$ is a characteristic sound velocity of the substrate. The frequency-dependent transmission probability between a semi-infinite wire (not finite beam) and a three-dimensional bulk solid similarly vanishes in the low frequency limit \cite{CrossPRB01,Chang&GellerPRB05}; however, that transmission probability cannot be directly used to estimate clamping loss because it would only be valid for beam lengths large compared to the wavelength (to be able to neglect the effect of cutting the wire to a finite length), and hence only to the higher frequency non-fundamental modes. We calculate the clamping $Q$ by regarding the resonator's  vibrational eigenfunction stress distribution (calculated with a fixed lower boundary) to be a source of elastic radiation into the substrate. The radiation problem is treated exactly within continuum elasticity theory, and the resonator vibrational modes are calculated numerically using finite-element methods. Our method applies to a wide variety of (but not necessarily all) NEMS resonator geometries, and can be used at high fundamental-mode frequencies where the subdominant terms in a multipole expansion would become important. 

The substrate is modeled as an isotropic elastic continuum, which is sufficient for the frequencies of interest. The substrate occupies $z > 0$, and at the surface $z=0$ the traction is ${\bf t}(x,y)$, which vanishes outside of the resonator contact region. The elastic displacement field ${\bf u}$ is decomposed into longitudinal parts and transverse parts ${\bf u}_{\rm l}$ and ${\bf u}_{\rm t}$, where
\begin{equation}
{\bf u}_{\rm l}({\bf r}) = \int \frac{d^2K}{(2 \pi)^2} \, {\bf U}_{\rm l}({\bf K}) \, e^{i {\bf K} \cdot {\bf r}} \, e^{i q_{\rm l} z}
\end{equation}
and
\begin{equation}
{\bf u}_{\rm t}({\bf r}) = \int \frac{d^2K}{(2 \pi)^2} \, {\bf U}_{\rm t}({\bf K}) \, e^{i {\bf K} \cdot {\bf r}} \, e^{i q_{\rm t} z}.
\end{equation}
Here 
\begin{equation}
q_{\rm l} \equiv \sqrt{\frac{\omega^2}{v_{\rm l}^2} - K^2} \ \ {\rm and} \ \ q_{\rm t} \equiv \sqrt{\frac{\omega^2}{v_{\rm t}^2} - K^2}.
\end{equation}
${\bf K} \equiv K_x {\bf e}_x + K_y {\bf e}_y$ is a two-dimensional wavevector. ${\bf U}_{\rm l}$ and ${\bf U}_{\rm t}$ satisfy $({\bf K} + q_{\rm l} {\bf e}_z) \times {\bf U}_{\rm l} = 0$ and $({\bf K} + q_{\rm t} {\bf e}_z) \cdot {\bf U}_{\rm t} = 0,$
and therefore can be written as ${\bf U}_{\rm l} = ({\bf K} + q_{\rm l} {\bf e}_z) \, \phi({\bf K})$ and
${\bf U}_{\rm t} =  {\bm  \Phi}({\bf K}) - q_{\rm t}^{-1} {\bf K} \cdot {\bm \Phi}({\bf K}) \, {\bf e}_z.$
Here $\phi$ is a scalar field, and ${\bf \Phi} \equiv \Phi_x {\bf e}_x + \Phi_y {\bf e}_y$ is a vector field. 

By construction, ${\bf u}_{\rm l}$ and ${\bf u}_{\rm t}$ satisfy the vector Helmholtz equations, conditions of longitudinality and transversality, and have only outwardly radiating components (for real $q$). The elastic ``potentials"  $\phi$ and ${\bm \Phi}$ are now fully determined by the three components of the surface traction 
\begin{equation} 
t_i = \lambda ({\bm \nabla} \cdot {\bf u}) \delta_{iz} + 2 \mu u_{iz}, 
\end{equation}
with $u_{ij}$ the strain tensor and $\lambda , \mu $ the Lam\'{e} coefficients. In terms of the Fourier transform ${\tilde {\bf t}}({\bf K})$ in the $xy$ plane we obtain
\begin{equation}
\begin{pmatrix} \Phi_x \\ \Phi_y \\ \phi \end{pmatrix} = S
\begin{pmatrix} {\tilde t}_x \\ {\tilde t}_y \\ {\tilde t}_z \end{pmatrix},
\end{equation}
where
\begin{widetext}
\begin{equation}
S \equiv \frac{i}{\chi} 
\begin{pmatrix} [K_x^2 (q_{\rm t}^2 - K_y^2) - (q_{\rm t}^2 - K_y^2)^2 - 4 K_y^2 q_{\rm l} q_{\rm t}]/q_{\rm t} &
K_x K_y ( K^2 - q_{\rm t}^2 + 4q_{\rm l} q_{\rm t})/q_{\rm t} & 2 K_x q_{\rm l} q_{\rm t} \\ 
K_x K_y ( K^2 - q_{\rm t}^2 + 4q_{\rm l} q_{\rm t})/q_{\rm t} & 
[K_y^2 (q_{\rm t}^2 - K_x^2) - (q_{\rm t}^2 - K_x^2)^2 - 4 K_x^2 q_{\rm l} q_{\rm t}]/q_{\rm t} &  
2 K_y q_{\rm l} q_{\rm t} \\ - 2 K_x q_{\rm t} &  - 2 K_y q_{\rm t} & K^2 - q_{\rm t}^2 \end{pmatrix}.
\end{equation}
Here $\chi \equiv  \mu [(K^2 - q_{\rm t}^2)^2 + 4 q_{\rm l} q_{\rm t} K^2].$ The elastic energy current, averaged over time, is 
\begin{equation}
j_z = \frac{\omega}{2} \, {\rm Im} \, ( \lambda \, u_z^* \, \nabla \cdot {\bf u} + \mu \, {\bf u}^* \cdot {\bf \nabla} u_z + \mu \, {\bf u}^* \cdot \partial_z {\bf u} ),
\end{equation}
and the radiated power is
\begin{equation}
P \equiv \int d^2r \, j_z = \frac{\omega}{2} \bigg[ \rho \omega^2 \! \int_{\Gamma_{\rm l}} \! \! \frac{d^2K}{(2 \pi)^2} \, \big| \phi \big|^2 q_{\rm l}
+ \rho v_{\rm t}^2 \! \int_{\Gamma_{\rm t}} \! \! \frac{d^2K}{(2 \pi)^2} \bigg( q_{\rm t} \big| {\bf \Phi} \big|^2 + \frac{1}{q_{\rm t}} \big| {\bf K} \cdot {\bf \Phi} \big|^2 \bigg) \bigg].
\label{power definition}
\end{equation}
\end{widetext}
The integration domains $\Gamma_{\rm l}$ and $\Gamma_{\rm t}$ in (\ref{power definition}) have $|{\bf K}|$ less than $\omega/v_{\rm l}$ and $\omega/v_{\rm t},$ respectively. Given the surface traction ${\bf t}(x,y)$, $P$ is now readily evaluated. The damping time $\tau$ is $E/P,$ with $E$ the initial energy of the resonator. The required traction information can be accurately determined by numerical finite-element methods, without the need to include a large ``environment" in the simulation. Our numerical results converge to the precision reported here after a few tens of thousands of volume elements.

\begin{table}
\caption{\label{Q table}
Resonator frequency $\omega/2\pi$ and $Q$ factor as a function of size $l$. Beams have a $5 \, {\rm nm}$ thickness and $10 \, {\rm nm}$ width.}
\begin{ruledtabular}
\begin{tabular}{|c|ccc|}
& $20 \, {\rm nm}$ & $50 \, {\rm nm}$ & $100 \, {\rm nm}$  \\ \hline
perpendicular & $14.7 \, {\rm GHz}$ &  $2.42 \, {\rm GHz}$ & $607 \, {\rm MHz}$ \\ 
beam               & $Q\!=\!4.1 \! \times \! 10^3$ &  $Q\!=\!2.5 \! \times \! 10^5$ & $Q\!=\!7.1 \! \times \! 10^6$ \\ \hline
cantilever       & $30.1 \, {\rm GHz}$ &  $3.20 \, {\rm GHz}$ & $695 \, {\rm MHz}$ \\ 
beam               & $Q\!=\!1.0 \! \times \! 10^3$ &  $Q\!=\!4.1 \! \times \! 10^5$ & $Q\!=\!1.4 \! \times \! 10^7$ \\ \hline
doubly clamped & $97.1\, {\rm GHz}$ &  $16.0 \, {\rm GHz}$ & $3.85 \, {\rm GHz}$ \\ 
beam               & $Q\!=\!21$ &  $Q\!=\!2.1 \! \times \! 10^3$ & $Q\!=\!7.9 \! \times \! 10^4$ 
\end{tabular}
\end{ruledtabular}
\end{table}

Our results are summarized in Table \ref{Q table}. The $Q$ factor for a perpendicular beam resonator shows that clamping loss in that case is probably negligible below $1 \, {\rm GHz}$, but that it becomes quite significant above $10 \, {\rm GHz}$. An approximate analytic expression for $Q$ in this geometry can be obtained by constructing the fundamental-mode eigenfunction with thin-beam theory, and keeping only the leading (monopole) transverse stress moment in (\ref{power definition}). This leads to \cite{dimensionsnote}
\begin{equation}
Q = \beta \, \bigg(\frac{w}{t}\bigg)^{\! \frac{1}{4}} \bigg(\frac{\Omega}{\omega}\bigg)^{\! \frac{5}{2}} \! \! ,
\label{monopole}
\end{equation}
where $\beta$ is a material-dependent parameter, equal to 0.194 for Si, and where
\begin{equation}
\Omega \equiv \pi v_{\rm l} A^{-1/2} \! , \ \ \ {\rm with} \ \ \  A \equiv t w.
\end{equation}
$\Omega$ is a characteristic frequency above which the resonator beam---with cross-sectional area $A$---can support nonuniform transverse modes, changing the dynamics from one dimensional to three dimensional. The validity of the result (\ref{monopole}) relies on the applicability of thin-beam theory; it is therefore asymptotically exact in the $\omega \ll \Omega$ limit. The beam-theory frequency and asymptotic $Q$ factor for the perpendicular resonator is given in Table \ref{perturbative Q table}, showing that the expression (\ref{monopole}) becomes unreliable (overestimating dissipation) above about $10 \, {\rm GHz},$ which is the regime where clamping loss just begins to be important. Also given in Table \ref{perturbative Q table} is a simpler (but not asymptotically exact) estimate by Photiadis and Judge \cite{PhotiadisAPL04},
\begin{equation}
Q_{\rm PJ} \approx 3.2 \, \frac{l^5}{w t^4},
\label{photiadis and judge}
\end{equation}
which predicts clamping loss rates somewhat higher than that given by (\ref{monopole}).

\begin{table}
\caption{\label{perturbative Q table}
Asymptotic Q factor (\ref{monopole}) for the perpendicular beam resonator. Dimensions are the same as in Table \ref{Q table}. The frequency is calculated analytically using thin-beam theory, and $\Omega/2\pi$ is $599 \, {\rm GHz}$. $Q_{\rm PJ}$ is taken from Ref.~[\onlinecite{PhotiadisAPL04}].}
\begin{ruledtabular}
\begin{tabular}{|c|ccc|}
perpendicular beam & $20 \, {\rm nm}$ & $50 \, {\rm nm}$ & $100 \, {\rm nm}$  \\ \hline
frequency & $17.1 \, {\rm GHz}$ &  $2.73 \, {\rm GHz}$ & $683 \, {\rm MHz}$ \\ 
asymptotic $Q$          & $2.5 \! \times \! 10^3$ &  $2.2 \! \times \! 10^5$ & $7.1 \! \times \! 10^6$ \\ 
$Q_{\rm PJ}$               & $1.7 \! \times \! 10^3$ &  $1.6 \! \times \! 10^5$ & $5.2 \! \times \! 10^6$ 
\end{tabular}
\end{ruledtabular}
\end{table}

The $Q$ factor for the cantilever and doubly clamped beam resonators behave quite similarly to the perpendicular resonator, when viewed as a function of frequency instead of size. {\sl In all cases the $Q$ factors at a few GHz are about $10^5$, but fall to around $10^3$ in the few tens of GHz regime}. Loss at $100 \, {\rm GHz}$ is severe. It is reasonable to assume that this trend is quite general, and will apply to other resonator geometries as well. The development of ultrahigh-$Q$ NEMS resonators above about $10 \, {\rm GHz}$ will require phononic bandgap engineering \cite{ClelandPRB01} or other methods of acoustic isolation.

Finally, we briefly comment on the high frequency behavior of the asymptotic expression (\ref{monopole}), which we have argued to be exact (for the perpendicular beam resonator) in the low-frequency limit. It follows from our earlier discussion that $Q$ would be expected to be of order unity at the frequency $\Omega.$ Thus, it is tempting to conclude that the formula (\ref{monopole}) can be used for all frequencies in the range $ 0 \lesssim \omega \lesssim\Omega$, which would be {\it incorrect}, despite the fact that it is accurate at the two endpoints.

This work was supported by the NSF under grants DMR-0093217 and CMS-040403.

\bibliography{/Users/mgeller/Papers/bibliographies/MRGpre,/Users/mgeller/Papers/bibliographies/MRGbooks,/Users/mgeller/Papers/bibliographies/MRGgroup,/Users/mgeller/Papers/bibliographies/MRGphonons,/Users/mgeller/Papers/bibliographies/MRGcm,/Users/mgeller/Papers/bibliographies/MRGqc-josephson,/Users/mgeller/Papers/bibliographies/MRGnano,clampingnotes}

\begin{thebibliography}{61}
\bibitem[{dim()}]{dimensionsnote}
\bibinfo{note}{Our notation for resonator dimensions is as follows: $l$ is the
  beam length, $t$ is the thickness in the flex direction, and $w$ is the width
  dimension transverse to that.}
\bibitem[{fun()}]{fundamentalnote}
\bibinfo{note}{Extension of our results to the lower frequency non-fundamental
  modes is straightforward, but for definiteness we focus on the fundamental
  mode.}
\end{thebibliography}


\end{document}